\documentclass[structabstract]{aa}  

\usepackage{natbib}
 \bibpunct{(}{)}{;}{a}{}{,} 
 
\usepackage{graphicx}

\usepackage{txfonts}
\def\kms{km s$^{-1}$\space}
\def\kmsnospace{km s$^{-1}$}
\def\kmsno{km s$^{-1}$}
\def\micron{$\mu$m\space}

\def\micronno{$\mu$m}
\def\arcsecno{$^{\prime\prime}$}
\def\arcsec{$^{\prime\prime}$\space}

\def\arcminno{$^{\prime}$}
\def\arcmin{$^{\prime}$\space}
\def\deg{$^{\circ}$\space}

\def\degno{$^{\circ}$}
\def\h2{H$_2$}

\def\ciis{[C\,{\sc ii}]}
\def\ciino{[C\,{\sc ii}]}
\def\hi{H\,{\sc i}\space}
\def\hino{H\,{\sc i}}
\def\hii{H\,{\sc ii}\space}

\def\cii{[C\,{\sc ii}]\space}
\def\13co{$^{13}$CO}
\def\c18o{C$^{18}$O}
\def\12co{$^{12}$CO}

\def\niinospace{[N\,{\sc ii}]}
\def\niino{[N\,{\sc ii}]}
\def\nii{[N\,{\sc ii}]\space}

\def\c+{C$^+$}

\def\h2{H$_2$}

\begin{document}

   \title{Ionized gas at the edge of the Central Molecular Zone}
   
\titlerunning{Ionized gas at the edge of the Central Molecular Zone}

\authorrunning{W. D. Langer et al.}

   \author{W. D. Langer
             \inst{1},
	P. F. Goldsmith
	 \inst{1},
              J. L. Pineda
              	 \inst{1},
              T. Velusamy
                  \inst{1},  
           M. A. Requena-Torres
            \inst{2}, 
            and
           H. Wiesemeyer
            \inst{2}
            }
      
         
   \institute{Jet Propulsion Laboratory, California Institute of Technology,
              4800 Oak Grove Drive, Pasadena, CA 91109, USA\\
              \email{William.Langer@jpl.nasa.gov}
	 \and                        
                  Max-Planck-Institut f$\ddot{\rm u}$r Radioastronomie, 
			Auf dem H$\ddot{\rm u}$gel 69, 53121 Bonn, Germany
             }

   \date{Received 18 November 2014; Accepted 15 January 2015}

 

\abstract
{The edge of the Central Molecular Zone (CMZ)  is the location where massive dense molecular clouds with large internal velocity dispersions transition to the surrounding more quiescent and lower CO emissivity region of the Galaxy. Little is known about the ionized gas surrounding the molecular clouds and in the transition region.}   {To determine the properties of the ionized gas at the edge of the CMZ near Sgr E using observations of N$^+$ and C$^+$.}  {We observed a small portion of the edge of the CMZ near Sgr E with spectrally resolved \cii 158\micron and \nii 205\micron fine structure lines at six positions with the GREAT instrument on SOFIA  and in \cii using \textit{Herschel} HIFI on-the-fly strip maps.  We use the \nii spectra along with a radiative transfer model to calculate the electron density of the gas and the \cii maps to illuminate the morphology of the ionized gas and model the column density of CO--dark H$_2$.} {We detect two \cii and \nii velocity components, one along the line of sight to a CO molecular cloud at -207 \kms associated with Sgr E and the other at -174 \kms  outside the edge of another CO cloud.  From the \nii emission we find that the average electron density is in the range of $\sim$ 5 to 21 cm$^{-3}$  for these features. This electron density is much higher than that of the disk's warm ionized medium, but is consistent with densities determined for bright diffuse \hii nebulae.  The column density of the CO--dark H$_2$ layer in the -207 \kms cloud is $\sim$ 1-2$\times$10$^{21}$ cm$^{-2}$ in agreement with theoretical models.  The CMZ extends further out in Galactic radius by $\sim$ 7 to 14 pc in ionized gas than it does in molecular gas traced by CO.}  {The edge of the CMZ likely contains dense hot ionized gas surrounding the neutral molecular material. The high fractional abundance of N$^+$ and high electron density require an intense EUV field with a photon flux of order 10$^6$ to 10$^7$ photons cm$^{-2}$ s$^{-1}$, and/or efficient proton charge exchange with nitrogen, at temperatures of order 10$^4$ K, and/or a large flux of X-rays.   Sgr E is a region of massive star formation as indicated by the presence of numerous compact \hii regions.  The massive stars are potential sources of the EUV radiation that ionize and heat the gas. In addition  X-ray sources and the diffuse X-ray emission in the CMZ are candidates for ionizing nitrogen.}

{} 

\maketitle



\section{Introduction}
\label{sec:introduction}

The Central Molecular Zone (CMZ), the central $\sim$400 pc by $\sim$ 80 pc of the  Galaxy, is  a region filled with  massive dense Giant Molecular Clouds (GMC) having large internal velocity dispersion.   Characteristic cloud line widths are $\ge$ 20 \kmsno, compared to a few \kms for GMCs in the  disk \citep[e.g.][]{morris1996,oka1998a,Ferriere2007}.  In addition, GMCs in the CMZ have higher kinetic temperatures,T$_k \sim$ 30 to 200 K \citep{Ao2013,Mills2013}, and densities, n(H$_2$) $\ge$ 10$^{4-5}$ cm$^{-3}$ \citep[e.g.][]{Jackson1996,oka1998a,Dame2001,Martin2004}, than GMCs in the  disk. The gas distribution within the inner Galaxy and across the edge of the CMZ is thought to be due to the gravitational potential and dynamics of the gas there \cite[see overview by][]{Ferriere2007}. While the atomic and molecular hydrogen distributions  in the CMZ have been mapped with spectrally resolved \hi and CO, respectively, less is known about the distribution and properties of the ionized gas in this region.  

The electron distribution of the interstellar medium has been determined from dispersion measurements of radio emission from pulsars and other compact radio sources \citep{Cordes2002,Cordes2003}.  \cite{Lazio1998}  and \cite{Roy2013} find that the average electron density in the inner 100 pc is $<$$n$(e)$> \sim$ 10 cm$^{-3}$. The radial distribution has been modeled as a Gaussian with scale length of 145 pc \citep[see][]{Ferriere2007}, which yields a value  at the edge of the CMZ $<n$(e)$> \sim$ 1 - 2  cm$^{-3}$. \cite{Lazio1998} proposed that the scattering of radio waves is not due to a uniform electron distribution but occurs in thin ionized layers at the surface of molecular clouds with $n$(e)$\gtrsim$10$^3$ cm$^{-3}$ and T$_{k}$(e) $\sim$ 10$^3$ K, or at the interface with the Hot Ionized Medium (HIM) with $n$(e)$\sim$ 5 -- 50 cm$^{-3}$ and T$_{k}$(e) $\sim$ 10$^5$ - 10$^6$ K.  Observations of the spectrally resolved emission from ions  in the CMZ can help distinguish among these different models and provide information about the density, filling factor, and ionization mechanisms. The objective of this paper is to explore the nature of the ionized gas at the edge of the CMZ  near Sgr E using spectrally resolved emission from the fine structure lines, \nii and \ciis, of ionized nitrogen, N$^+$, and ionized carbon, C$^+$, respectively. 

The Balloon-borne Infrared Carbon Explorer (BICE) mapped \cii in the inner Galaxy with modest spatial (15\arcminno)  and low spectral ($\sim$175 \kmsno) resolution \citep{Nakagawa1998}. BICE detected bright \cii emission from the CMZ from 358\fdg 3 $< l <$  1\deg but tapering off outside, and confined to $|b| \lesssim$1\deg (see their Figures 7 and 8).  Within the limitations of BICE's low spectral resolution, \cite{Nakagawa1998}  found that \cii has a large velocity dispersion towards the CMZ, $\Delta$v$\ge$100 \kms (see their Figure 14), similar to the dispersion seen in $^{12}$CO.  In addition, the Infrared Space Observatory (ISO) observed \cii and \nii along a number of lines of sight in the CMZ  \cite[see][]{Rodriguez2004} with a spectral resolution $\sim$ 30 \kmsno. However, much better spectral resolution is needed to separate  individual components and determine gas characteristics.  

To understand better the distribution of the ionized gas at the edge of the CMZ  
we observed several lines of sight in  \nii and \cii in July 2013 using the heterodyne  German REceiver for Astronomy at Terahertz frequencies  \cite[GREAT\footnote{GREAT is a development by the MPI f$\ddot{\rm u}$r Radioastronomie and KOSMA/Universit$\ddot{\rm a}$t zu K$\ddot{\rm o}$ln, in cooperation with the MPI f$\ddot{\rm u}$r Sonnensystemforschung and the DLR Institut f$\ddot{\rm u}$r Planentenforschung.};][]{Heyminck2012} onboard the Stratospheric Observatory for Infrared Astronomy (SOFIA) \citep{Young2012}. In addition, prior to the GREAT observations, we made small scale strip maps in \cii to map the overall distribution of ionized gas as part of a {\it Herschel}\footnote{{\textit Herschel} is an ESA space observatory with science instruments provided by European-led Principal Investigator consortia and with important participation from NASA.} \citep{pilbratt2010} Open Time 2 Programme, using the high spectral resolution Heterodyne Instrument in the Far Infrared \cite[HIFI;][]{degraauw2010}.  

Here we report on the results of these observations, which indicate that there are two regions of highly ionized, high temperature, dense gas: one along the line of sight that includes a CO molecular cloud associated with Sgr E at V$_{lsr}\sim$ -207 \kms and the other at $V_{lsr}\sim$ -174 \kms outside the edge of a CO molecular cloud located at larger $l$ in the CMZ. The electron densities derived from the \nii emission, $n$(e) $\sim$ 5 to 25  cm$^{-3}$, are two to three orders of magnitude higher than that in the disk's  warm ionized medium (WIM), but consistent with those found from analysis of \nii emission in the Carina Nebula, a luminous \hii region with numerous O stars \citep{Oberst2011}. This paper is organized as follows.  In Section 2 we present our observations, while Section 3 analyzes the distribution and density of the ionized gas.  Section 4 discusses the possible ionization sources, and Section 5 summarizes the results.



\section{Observations of \cii and \nii}
\label{sec:observations}

\begin{figure}
 \centering
     \includegraphics[width=9cm]{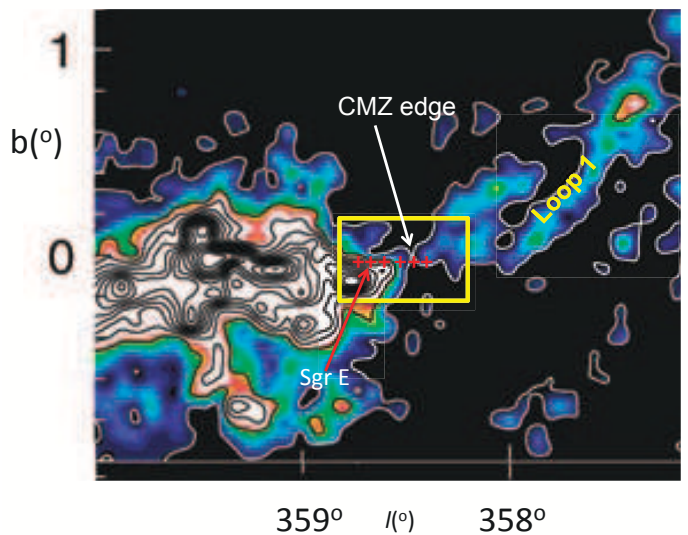}
      \ \caption{A portion of a map of CO integrated intensity from -180 to -90 \kms of the edge of the CMZ from \citep{fukui2006}, and reprinted with permission from AAAS, based on the NANTEN CO(1-0) map which has a beam size $\sim$3\arcminno. The region mapped with HIFI is outlined by the yellow box and the positions of the \cii and \nii observations with GREAT are indicated by crosses.  The approximate  CO edge of the CMZ and a portion of Loop 1 are indicated, as is the location of Sgr E.} 
         \label{fig:fig1}            
 \end{figure}

In Figure~\ref{fig:fig1} we show a portion of the NANTEN 4-m CO map from Figure 1a in \cite{fukui2006}, and reprinted with permission from AAAS,  that zooms in on the CO edge of the CMZ.  The yellow box in Figure~\ref{fig:fig1} outlines the region mapped with HIFI, and the positions observed in \cii and \nii with GREAT are indicated by red crosses.   \nii traces just the highly ionized gas, while \cii  traces highly ionized gas and the weakly ionized neutral atomic and molecular gas.  Located just inside the edge of the CMZ is Sgr E, a region with many radio sources  located around  $l \sim$ 358\fdg 7 and $b \sim$ 0\deg \citep{Gray1994}, which contains 18 compact ($<$2\arcminno) \hii regions and diffuse emission  \citep{Liszt1992}.  Sgr E is a region of active star formation associated with a CO molecular cloud with V$_{lsr} \sim$ -210 \kms \citep{oka1998a}. The radio recombination line survey, \hii Region Discovery Survey \cite[HRDS:][]{Anderson2011,Anderson2012} detected eight of these \hii sources towards Sgr E at $V_{lsr} \sim$ -208 \kmsno, the same velocity as the CO cloud.  Along the line of sight outside  the CMZ there are Giant Magnetic Loops, large scale molecular gas  loop-like features, with large line widths and large scale flows detected in CO \citep{fukui2006,Fujishita2009}.   The edge of Loop 1 intercepts the plane at $l \sim$358\fdg 2 to 358\fdg 4, close to the line of sight to the edge of the CMZ, and has a  V$_{\rm lsr} \sim$ -135 \kms \citep{fukui2006}.  This loop, however, is at a Galactic radius, R$_G \sim$670 pc \citep{torii2010a}, well outside the CMZ. 
 
\subsection{SOFIA \cii and \nii data}

We observed the lower fine structure transition of ionized nitrogen, \nii $^3P_1$$\rightarrow$$^3P_0$ at 1461.1319 GHz (205.2 $\mu$m) and, simultaneously, the fine structure line of ionized carbon, \cii $^2P_{3/2}$$\rightarrow$$^2P_{1/2}$ at 1900.5369 GHz (157.7 \micronno), using GREAT.  The  angular resolution is 15\arcsec at \cii and 20\arcsec at \niino.  To improve the signal-to-noise ratio (SNR) the \cii and \nii data were smoothed to a velocity resolution of 1.20 \kms and 1.57 \kms per channel, respectively.   The spectra were calibrated using the procedure outlined in \cite{Guan2012}. 

 The GREAT  \cii observations  contained atmospheric features in the spectral region observed due to the frequency setting for large negative velocities of the source, and the \cii line was in the wing of a strong water line.  Moreover, the position switching observing mode made the presence of this feature a significant issue that needed to be addressed in the data reduction.  The standard atmospheric calibration  was not sufficient to produce a satisfactory fit to the water feature, and we had to use a set of non-standard parameters in the atmospheric model \citep{Guan2012} to fit the width and the peak of the water line. The main change to the normal procedure was to reduce the contribution from the Dry atmosphere in steps, allowing the Wet part (water) to fit the atmospheric feature. We implemented a set of six models, and chose the model that best fit the water feature. This feature was removed and a zero order spectral baseline fit applied to the delivered data products. The resulting spectra were, with one exception, fairly regular with residual low order polynomial variations and only slight offsets from zero; one spectrum had a long period, but well ordered, variation in the baseline.   In order to fit the baselines better than zero order, the  delivered data were then reduced further in CLASS\footnote{http://www.iram.fr/IRAMFR/GILDAS} and baselines corrected with a polynomial fitting routine. Comparison of the GREAT and HIFI \cii data at one nearly coincident position (see below) shows excellent agreement, lending confidence to the removal of the water line and overall baseline fittings.  The intensities were corrected for a main beam efficiency of 0.67.  The rms noise for the final \cii products ranges from 0.16 to 0.26 K per 1.20 \kms channel, while that for \nii ranges from 0.09 to 0.23 K per 1.57 \kms channel.   

\begin{figure}
 \centering
      \includegraphics[width=8cm]{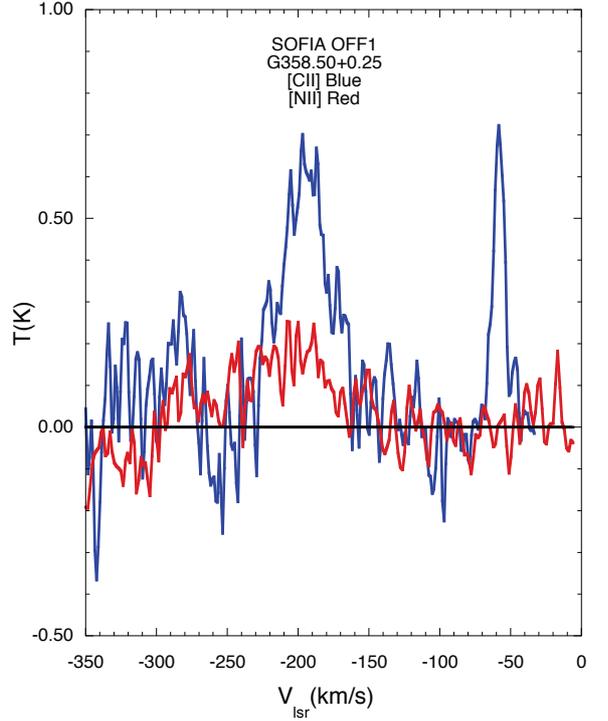}
      \caption{\cii and \nii emission in the reference OFF1 position at $l$=358\fdg 50 and $b$=0\fdg 25. The data have been smoothed to 2.4 \kms for \cii and 3.1 \kms for \nii to improve the signal-to-noise. The \cii emission peaks at V$_{lsr} \sim$ -195 \kms while that from \nii peaks $\sim$ -207 \kmsno, within the fitting errors of the weaker \nii signal.  The narrow \cii emission at V$_{lsr} \sim$ -58 \kms arises from the 3 kpc arm.  \nii is not detected from this arm.}
           \label{fig:fig2}
 \end{figure}

We observed six lines of sight (LOS) along $b$=0\deg at longitudes from $l$=358\fdg 45 to 358\fdg 75.  The reference off-position, OFF1, $l$=358\fdg 50 and $b$=0\fdg 25 contains weak \cii and very weak \nii emission, shown in Figure~\ref{fig:fig2}, so that all the on-positions had to be corrected for this emission.  To correct the observations for emission in OFF1 it was observed separately on two nights as an on-position using a different off-position, OFF2, $l$=358\fdg 50 and $b$=0\fdg 65. The OFF1 lines appear to be singly peaked and  a Gaussian fit to the \cii emission in OFF1 peaks at V$_{lsr} \sim$ -195 \kms and has a full width half maximum (FWHM) $\sim$ 43 \kmsno.  The narrow \cii feature at V$_{lsr} \sim$ -58 \kms (FWHM $\sim$ 9 \kmsno) arises from the 3 kpc arm.  A Gaussian fit to the broad, weak  \nii line yields a peak V$_{lsr} \sim$ -205 \kms and a FWHM $\sim$65 \kmsno, close to the characteristics of the \cii line considering the weakness of the line and the low signal-to-noise ratio of this detection.   No other \nii is detected in the band above the 3-$\sigma$ level.  

\begin{figure}
 \centering
      \includegraphics[width=8cm]{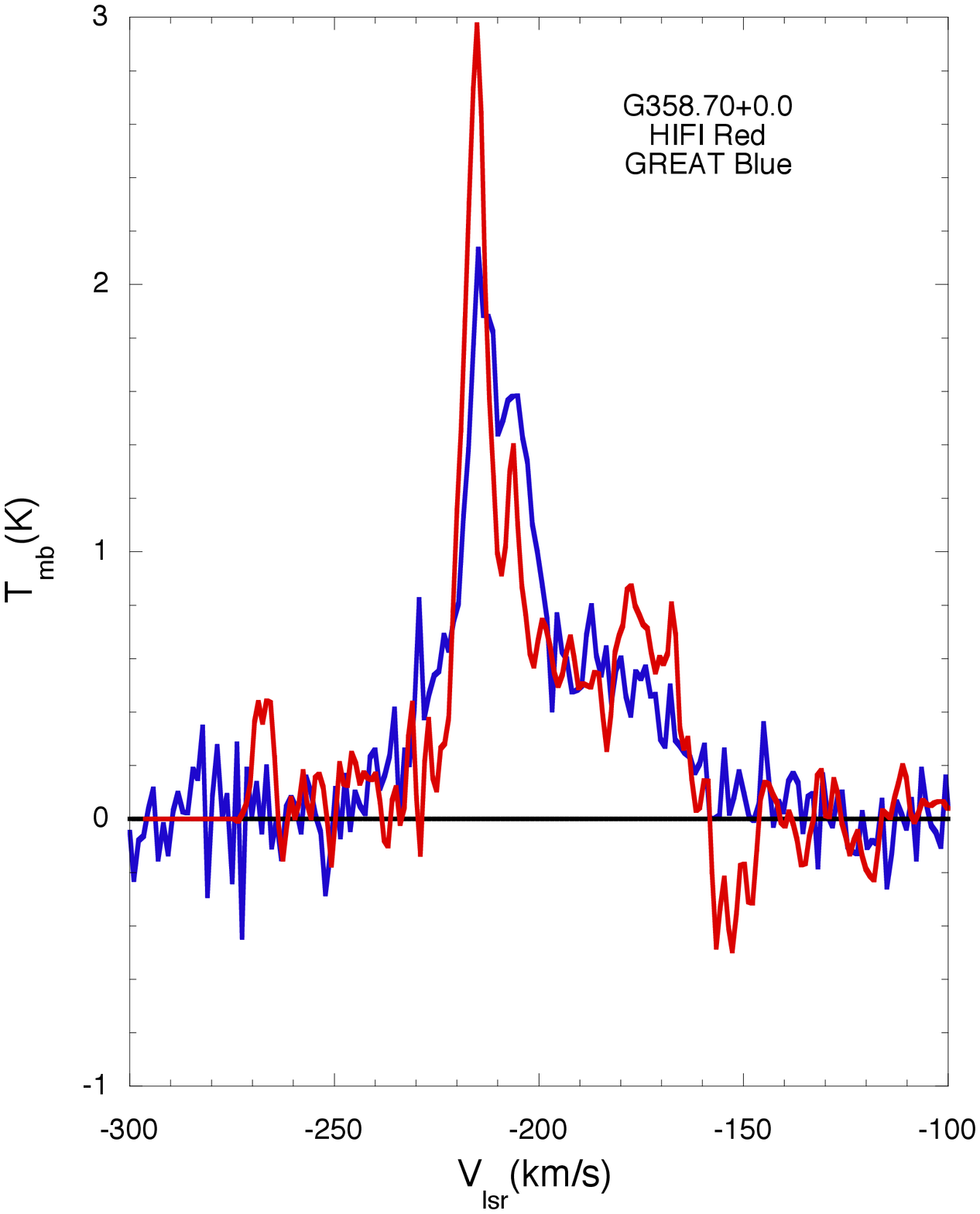}
      \caption{GREAT \cii spectrum at the position observed with GREAT, $l$=358\fdg700 and $b$=0\fdg0, that is closest to a \cii spectrum  from GOT C+.  The GOT C+ position $l=$358\fdg 696 and $b=$0\fdg 0 is offset in longitude by 14\arcsec from the GREAT pointing, which is about the GREAT beam width. The HIFI data have been smoothed to the resolution of the GREAT data, 1.2 \kmsno.  The \cii GREAT spectrum is similar to that from HIFI showing both the -207 \kms and -174 \kms components.  The total integrated intensities agree within $\pm$10\%.}
               \label{fig:fig3}
 \end{figure}

As a check on the quality of the GREAT data, given that it had to be corrected for the atmospheric water line and emission in the OFF1 position, we show in Figure~\ref{fig:fig3} a  spectrum from the HIFI Galactic Observations of Terahertz (GOT C+) \cii survey \citep{Langer2010,Pineda2013} at the closest position, $l=$358\fdg 696 and $b=$0\fdg 0, to the GREAT observation at $l=$358\fdg 700 and $b=$0\fdg 0.  The offset in longitude of 14\arcsec corresponds to 1.2 HIFI and 0.9 GREAT beam widths, and some portions of the main beams are only about an arcsecond apart.    The \cii emission from HIFI and GREAT  have the same distribution in velocity, both show a strong peak at $\sim$ -210 \kms with intensities that generally agree within the uncertainties  of the baseline fitting, and both show weak emission centered around  $\sim$ -175 \kms that blends into the stronger feature. The integrated intensities of these two observations over the velocity range -150 \kms to -250 \kms differ by only 10\%. Furthermore, the GOT C+ spectrum used a different off position than the GREAT observations, lending further support for our procedure to correct the GREAT  \cii and \nii for  emission in the GREAT OFF1 position.  

\begin{figure*}[!ht]
\centering
         \includegraphics[width=16cm]{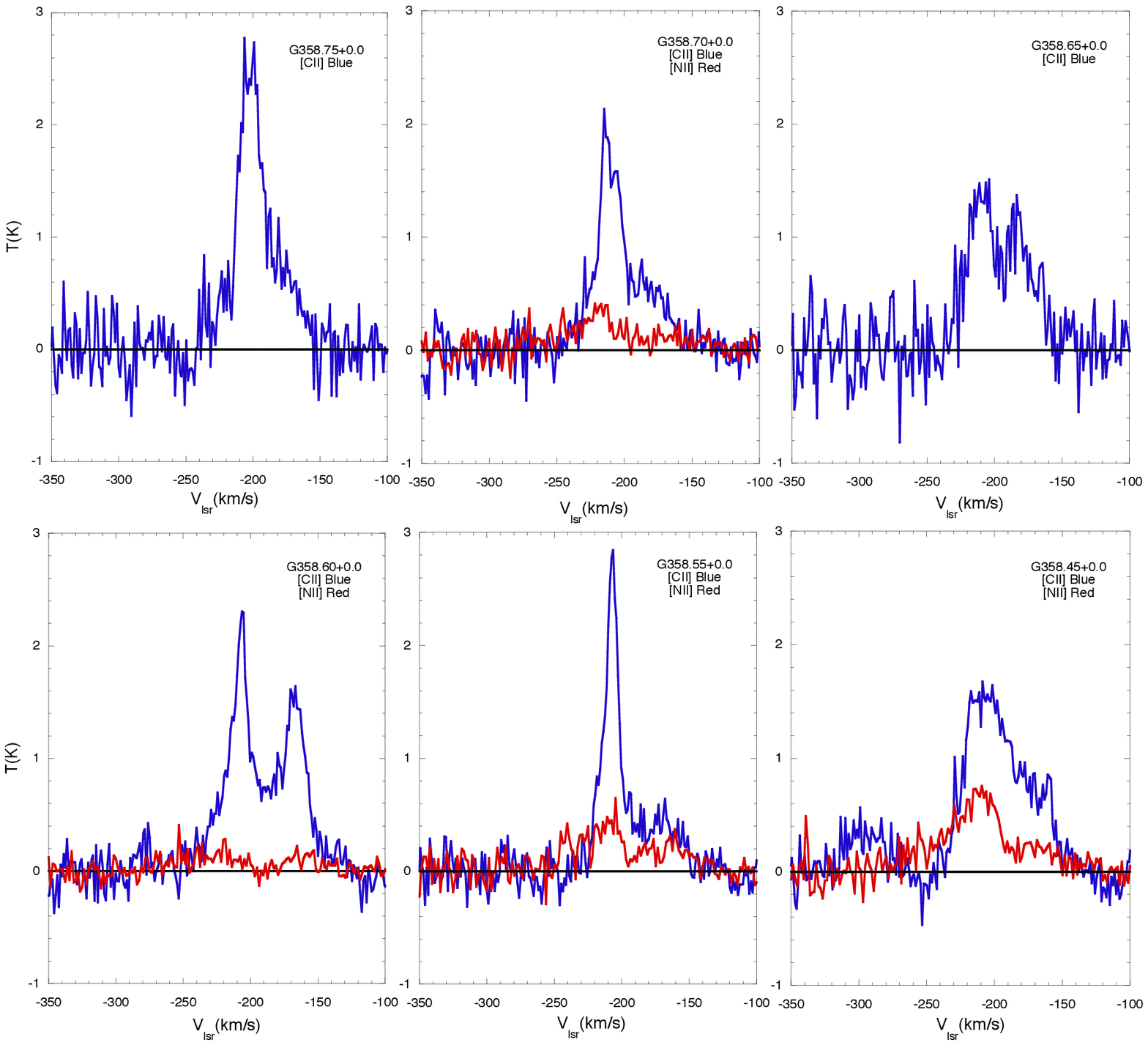}
      \caption{\cii (blue) and \nii (red) spectra at the six positions observed with GREAT along $b$=0\fdg0. The Galactic longitude and latitude are indicated by GXXX.XX+Y.Y.  The two features centered at $\sim$ -174 \kms and $\sim$ -207 \kmsno (which are blended at some positions)  arise from the edge of the CMZ.  \cii is detected at all 6 positions in the CMZ, but \nii is only detected at 4 positions.  The two non-detections in \nii are associated with greater rms noise.  While \nii is generally much weaker than \ciino, the \nii peak temperature at $l$=358\fdg 45 is about half that of \ciino.}
              \label{fig:fig4}
        \end{figure*}

The \cii and \nii spectra along $b$=0\deg are shown in Figure~\ref{fig:fig4}.  \cii emission was detected at all six positions and there are two adjacent, and at times blended, features arising from the CMZ in the velocity range $\sim$-150 \kms to -250 \kmsnospace.    At the two positions where these components are clearly separated, the peaks occur at V$_{lsr} \sim$ -174 and -207 \kmsno.  In general the -174 \kms component is weaker than the -207 \kms component, and both features are broad with a typical FWHM of $\sim$ 25 \kmsno. No emission is detected at  V$_{lsr}$$\sim$ -135 \kmsno, the velocity of Loop 1.  

\nii is only detected at four of the six LOS, and is generally much weaker than \cii except at $l$=358\fdg 45 where the component at V$_{lsr} \sim$ -207 \kms is about half as intense as \ciino. The four LOS where \nii is detected have an rms noise of $\sim$0.1 K per 1.57 \kms  channel whereas the two lines of sight where \nii is not detected have a larger rms noise, $\sim$0.23 K.  The  \nii components were fit with a double Gaussian with one peak at $\sim$ -215 \kms and  a FWHM $\sim$35 \kms and the other component is at $\sim$ -170 \kms with a FWHM $\sim$25 \kmsno.  All the \cii features have a large SNR while the \nii SNR ranges from $\sim$4 to 20 for the four LOS in which the line was detected.

\subsection{Herschel \cii map data}

The GREAT observations provide \cii at six and \nii at four longitudinal positions in the plane  ($b =$ 0\degno). To get insight into the distribution of \cii above and below the plane we use \cii maps made with the high spectral resolution HIFI \citep{degraauw2010} instrument onboard {\it Herschel} \citep{pilbratt2010}. We used a number of On-The-Fly (OTF) scans to map the 0\fdg 6 $\times$ 0\fdg 4 region shown in Figure~\ref{fig:fig1}. All HIFI OTF scans were made in the LOAD-CHOP mode using a reference off-source position at $l$=358\fdg 80, $b$=1\fdg 90, about 1\fdg 7 away in latitude from the top of the map. The map consists of 13 scans in Galactic latitude, centered at $b$ = 0\deg and stepping in longitude at 3\arcmin (0\fdg 05) intervals from 358\fdg 20 to 358\fdg 80.  The OTF $b$-scans are 24\arcmin long, sampled at every 40 \arcsecno.    At 1.9 THz the angular resolution of the Herschel telescope is 12\arcsecno, corresponding to a spatial resolution of $\sim$0.5 pc at 8.5 kpc, the distance to the Galactic Center. However the effects of  scanning rate and the 40\arcsec sampling of the OTF maps degrades the effective resolution in $b$ to 80\arcsec while it is 12\arcsec in $l$. Thus spatial structures on scale sizes smaller than 80\arcsec in $b$ will be strongly attenuated. Furthermore the total duration of each OTF scan was typically $\sim$2000 sec which provides only a small integration time on each spectrum (pixel).  Thus the rms in the OTF spectra, 0.22 K in T$_{\rm mb}$ in an 88\arcsec beam and 3 \kms channel,  is larger than that in the corresponding GREAT spectra. 

\begin{figure}
 \centering
   \includegraphics[width=9.0cm]{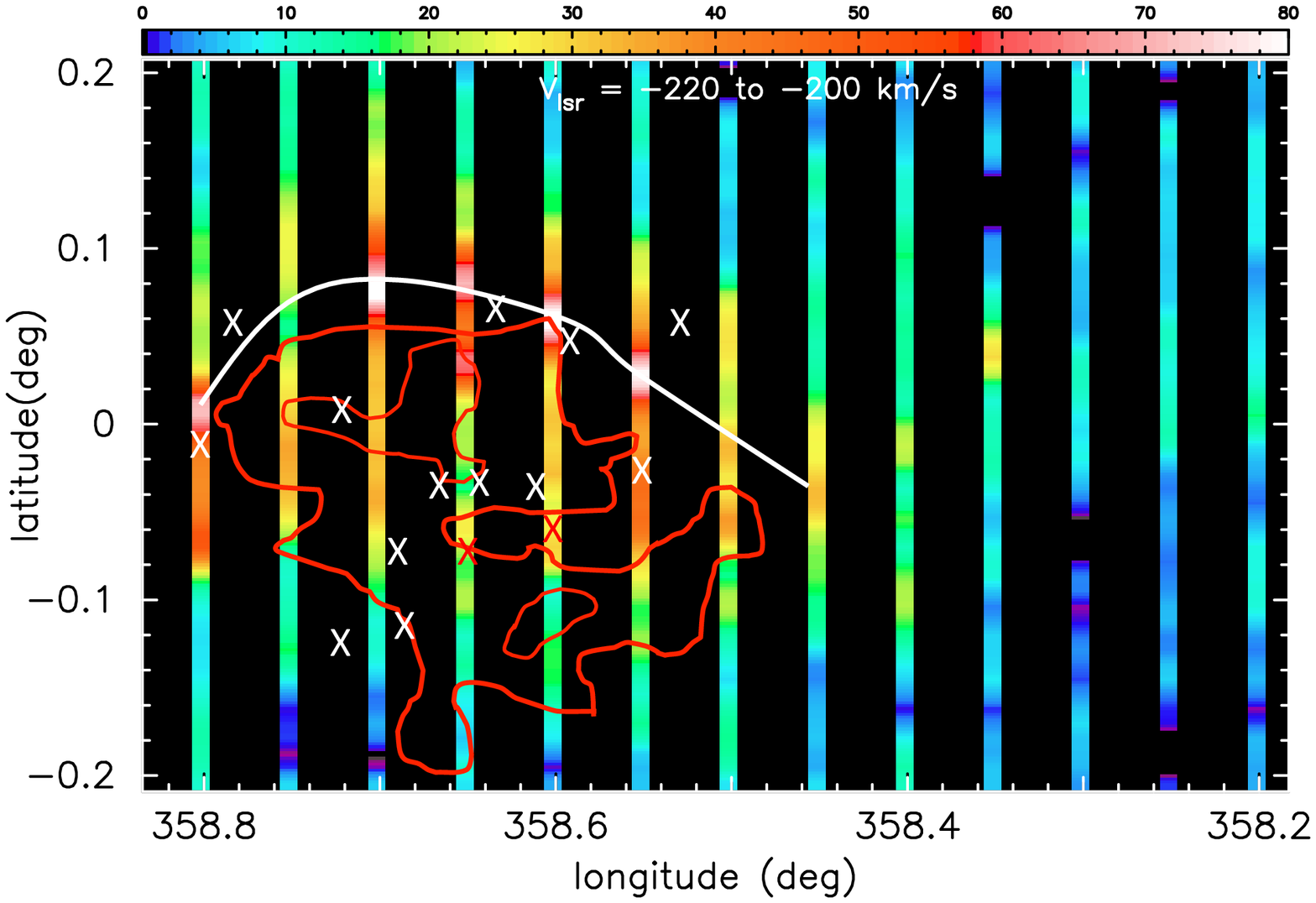}
      \caption{Comparison of the \cii OTF intensity maps  integrated over the velocity range -220 \kms to -200 \kms shown as coulored strips with the CO map from \cite{oka1998a} integrated over the velocity range -220 to -210 \kmsno.  The CO integrated intensity maps from \cite{oka1998a} are shown only in 10 \kms steps, but to improve the signal to noise in the \cii OTF maps we use a range of 20 \kmsno. The CO map integrated from -210 to -200 \kms shows the same extent in $b$ and the same edge at $l \sim$ 358\fdg 5 as the -220 to -210 \kms map, but extends further to 359\fdg 0, and would give the same morphological relationship to the \cii contours. The color bar at the top indicates the \cii integrated intensity in K \kmsno. The \cii OTF maps sample in $b$ every 0\fdg 05 in longitude from 358\fdg 2 to 358\fdg 8, so generating a contour map is impractical due to the sparse sampling in $l$.  Instead we have indicated by the white line the connection of the peak intensity in each strip.  The CO contours, indicated in red, are in steps of 15 K \kmsno.  The \cii is strongest slightly beyond the edge of the CO emission above the plane.  The compact \hii sources towards Sgr E  \citep{Liszt1992} are indicated by white and red X symbols, depending on background colour.}
              \label{fig:fig5}
 \end{figure}

The OTF map data presented here were processed following one of the procedures suggested by the Herschel Science Center science staff, by running the HIPE pipeline without an off-source subtraction to produce level 1 data.  Then the {\it hebCorrection} routine was applied to the level 1 data to remove the HEB standing waves and afterwards we followed the procedure described in \cite{Velusamy2014} to produce the level 2 data and spectral line maps.  The fact that {\it hebCorrection} subtracts the matching standing wave patterns from  a large database of spectra eliminates the need for off-source subtraction. Thus in our analysis the processed spectral data are free from any off source contamination.  We use the processed spectral line scan map data to make latitude--velocity ($b$--V) maps of the velocity structure of the \cii emission as a function of latitude at each of the 13 longitudes of the OTF positions. To increase the sensitivity of the $b$--V maps we use a velocity resolution of 3 \kmsno. In all the OTF $b$--V maps the velocity feature near -207 \kms is prominent, but that of the weaker -174 \kms feature is less so, possibly due to the combined effect of low sensitivity and beam dilution. However, the bright \cii features ($>$ 1K) are well traced in the OTF maps.  To bring out the spatial structure of the -207 \kms velocity feature we generate a \cii integrated intensity map covering -220 \kms to -200 \kmsno. This velocity range facilitates comparison with the Nobayama CO maps of \cite{oka1998a}, which are presented in 10 \kms intervals. To generate the \cii map we integrate the intensities in the $b$--V maps at each latitude pixel over the range of velocities from -220 \kms to -200 \kms around the \cii spectral peak. We then use these intensities at all longitudes to generate the 2--D map shown in Figure~\ref{fig:fig5}. Note that while the intensities are sampled contiguously in latitude, we only have data at 13 discrete longitudes with a pixel width of 12\arcsecno. To make the data more visible on the map we broadened their display in longitude to 60\arcsecno. There is little or no \cii for $l <$ 358\fdg 30.

\section{Results}
\label{sec:results}

The relationship between the distribution in \cii and CO is brought out in Figure~\ref{fig:fig5} where we superimpose on the \cii strip maps the CO integrated intensity from -210 \kms to -220 \kms \citep{oka1998a}.  The only caveat is that the region at the edge of the CMZ is irregularly mapped in CO in the Nobayama maps.  However, the AT\&T Bell Labs CO survey mapped the CMZ out to $l = \pm$2\deg \cite[see Figure 1;][]{morris1996} and also shows a sharp edge in CO emission at $l \sim$358\fdg 5. The -207 \kms cloud is clearly seen in the position-velocity maps from the AT\&T Bells Labs CO survey \cite[see Figure 4;][]{morris1996}. The CO emission comes from the cloud associated with Sgr E, an active region of star formation with bright compact \hii regions. Though our \cii map is not contiguous in longitude (scans are spaced by 3\arcminno) the brightest emission (white in the color wedge) appears to be an arc above the plane slightly outside the lowest CO contour.  The \cii arc is most likely limb brightened emission.  There does not appear to be a similar limb brightening for $b <$0\degno. Although we do not have any \nii observations along this arc, we conclude from the presence of \nii along the line of sight to the CO cloud at $b$ = 0\deg that  the \nii and  \cii emission arise from an ionized region surrounding the cloud, but there is also a likely contribution to \cii from a photon dominated region (PDR) sandwiched between the highly ionized hydrogen and dense molecular gas traced by CO. Also indicated on the map are 15 of the 18 compact \hii sources detected by \cite{Liszt1992} towards Sgr E (the remaining three are outside the mapped region). In addition to these compact \hii regions there appears to be low level diffuse emission from $l \sim$ 358\fdg 8 to 359\fdg 0 between $b = \pm$0\fdg1 (see Figure 8 in \cite{Liszt1992}). The compact \hii sources have electron densities of order several hundred cm$^{-3}$ and require ionizing stars of type B0 or brighter \citep{Liszt1992}.

All six \cii and the four \nii spectra in Figure~\ref{fig:fig4} show emission in the velocity range -140 to -240 \kmsno.  The detection of broad emission in \cii and \nii indicates the presence of warm, and perhaps dense, ionized gas at the edge of the  CMZ.  To gain further insight into the nature of the gas and its association with different interstellar medium (ISM) components we compare \cii and \nii to the CO(1-0) and \hi spectra along $b=$ 0\deg taken from the Three-mm Ultimate Mopra Milky Way Survey\footnote{www.astro.ufl.edu/thrumms} (ThrUMMS) \citep{Barnes2011} and the Australia Telescope Compact Array \hi Galactic Center survey\footnote{www.atnf.csiro.au/research/HI/sgps/GalacticCenter} \citep{McClure2012}, respectively.    

\begin{figure}
 \centering
   \includegraphics[width=8cm]{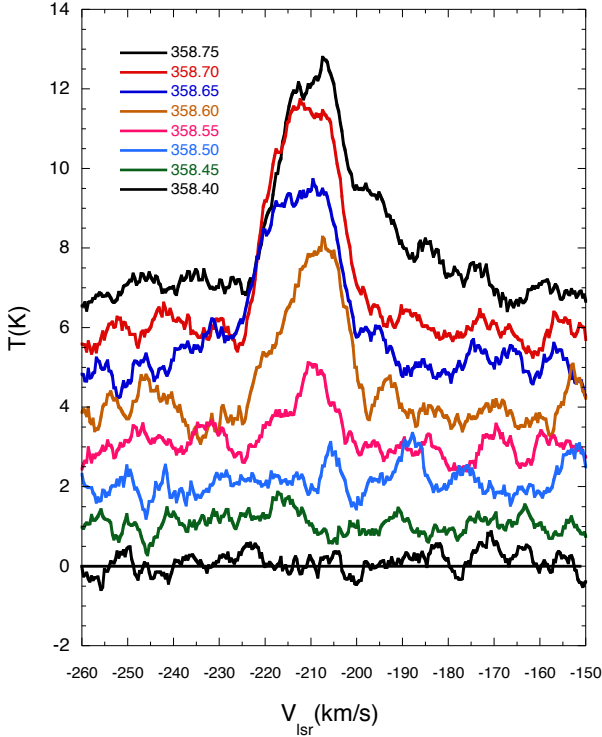}
       \caption{CO(1-0) spectra at $b=$ 0\deg at the longitudes observed in \cii and \niino. The data is taken from the ThrUMMS survey  (uncorrected for main beam efficiency)  and the inset is a key to the lines of sight observed. Moving towards the CMZ, larger $l$, CO is not detected until $l \sim$ 358\fdg 55 (although there is a hint of CO at 358\fdg 50) at the inner edge of the CMZ, and that only the -207 \kms component is present.  The -174 \kms CO component is detected at 358\fdg 75, the position furthest from the edge of the CMZ and it is considerably weaker there than the -207 \kms component. }
         \label{fig:fig6}       
 \end{figure}

\begin{figure}
 \centering
  \includegraphics[width=8cm]{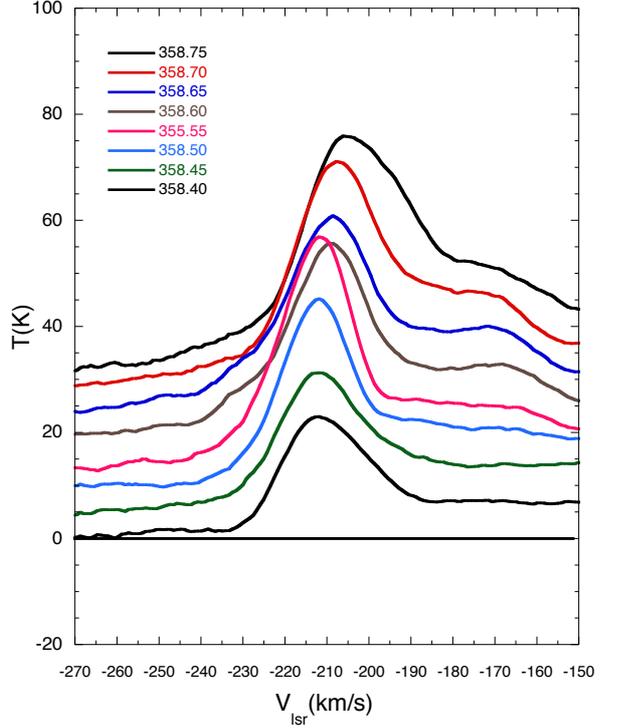}
      \caption{\hi spectra at $b=$ 0\deg along the latitudes observed in \cii and \niino. The data are taken from the ATCA \hi survey \citep{McClure2012} and the inset is a key to the lines of sight observed.  \hi is detected at all positions at -207 \kms and is strongest close to Sgr E.  \hi is also detected at  -174 \kms but is weaker than at -207 \kmsno. }                            
       \label{fig:fig7}    
 \end{figure}

In Figure~\ref{fig:fig6} we plot the CO(1-0) spectra at $b=$ 0\deg along the strip observed with GREAT, $l =$ 358\fdg 40 to 358\fdg 75 and $b=$ 0\degno. The corresponding \hi spectra are plotted in Figure~\ref{fig:fig7}. The CO and  \hi spectra show that there are two different environments detected in \cii and \niino. The emission at V$_{lsr} \sim$ -207 \kms is associated with both CO and \hino, while that for -174 \kms is associated only with \hi until $l \sim$ 358\fdg 75, where there is a hint of CO emission.  The distributions of the \cii and \nii emission for the -207 \kms component are towards the edge of a CO molecular cloud associated with Sgr E, while the  -174 \kms distribution is along the edge of another CO molecular cloud, as can be seen in the Nobayama CO maps in this velocity range \citep{oka1998a}.

To evaluate the properties of the ionized gas  we derived the parameters of the \ciino, \niino, and CO from Gaussian fits to each line component.  In cases where the  lines are blended we derive their line parameters using a double Gaussian fit over the velocity range -100 \kms to -300 \kmsno.  The intensities are given in Table~\ref{tab:Table1}.  We calculate the corresponding \hi intensities by integrating over the velocity ranges of the two \cii components because we cannot get unambiguous Gaussian fits over such  broad and, at low velocities, flat lines.  To avoid double counting the intensities we chose -190 \kms as the dividing velocity of the two components, and their intensities are given in Table~\ref{tab:Table1}.  In general, there is a good correlation of  \nii intensity with \cii intensity, which  suggests that a significant fraction of the \cii emission arises in the \nii region.   We will use the \cii and \nii integrated intensities to characterize the properties of the ionized gas. 
 
\begin{table}[!htbp]																	
\caption{Integrated line intensities}
\label{tab:Table1}
\label{tab:SOFIA_Lines}															
		\begin{tabular}{lcccc}																	
\hline	
  LOS &  I(\ciis)$^{a,b}$  &  I(\niinospace)$^c$  & I(CO) & I(\hino) \\
    \hline
  \hline 
V$_{lsr}$ = -207    \kms &  &  \\ 
\hline
358.45+0.0  & 63.4 & 26.4  & 6.7&  1047 \\
358.55+0.0  & 45.4 & 15.4  & 26.6 & 1904 \\
358.60+0.0  & 47.6 & 7.1 & 60.6 & 2189 \\
358.65+0.0  &38.6  &  -  & 101.9 & 2539 \\
358.70+0.0  &45.4  & 12.4  & 98.9 &  2923 \\
358.75+0.0  & 57.4 &  -  & 110.5 & 3235  \\
\hline
V$_{lsr}$ = -174    \kms &   &   & &  \\ 
\hline
358.45+0.0  & 21.1 & 8.2 & - & 730  \\
358.55+0.0  & 12.2 & 8.8  & - & 1212 \\
358.60+0.0  &  43.0 & 3.7  &  - &  1544\\
358.65+0.0  & 25.5 &  -  & - &  1855 \\
358.70+0.0  & 15.3  & 5.3  & - &  2177 \\
358.75+0.0  & 21.8 &  - & 9.8  & 2522  \\
\hline	
\end{tabular}
\\	
a) Integrated intensities are in units of K \kmsno.  We only report detections with a SNR $\ge$ 3, see text.  b) Typical rms noise in the \cii  integrated intensity is $\sim$1.4 K \kmsno. c) Typical rms noise in the \nii ntegrated intensity is $\sim$1.3 K \kmsno.  
\end{table}

\subsection{Radiative transfer solutions for \cii and \nii}

The relationship of the column density of C$^+$ and line intensity, I(\ciino), has been extensively discussed in the literature \cite[e.g.][]{Goldsmith2012}. The \cii peak antenna temperatures reported here are $<<$0.3 times the kinetic temperature and, as discussed by  \cite{Goldsmith2012}, are in the ``effectively optically thin limit" in which the column density is proportional to the integrated line intensity.  Therefore we have from \cite{Langer2014}, Equation A.2, the relationship between the column density, $N$(C$^+$) in cm$^{-2}$ and integrated line intensity, $\int T(K)d\rm{v}$, in K \kmsno,

\begin{equation}
N(C^+)=2.92\times10^{15}[1+0.5e^{\Delta E/kT}(1+\frac{n_{cr}}{n})]I([C\,II])\,\,\rm{cm^{-2}}
\end{equation}  

\noindent where $\Delta E/k$ =91.25K is the energy needed to excite the $^2P_{3/2}$ level, $n$ is the density of the collision partner of the bulk of the gas, and $n_{cr}$ is the corresponding  critical density.  The collision rate coefficients for e, H, and H$_2$ are summarized in \cite{Goldsmith2012} with an update for H$_2$ in \cite{Wiesenfeld2014}.  Electrons will dominate the excitation of C$^+$ where we detect \niino.  In the  ionized regions where \nii is detected the temperatures are high enough that we can neglect the exponential term and

\begin{equation}
N(C^+)=2.92\times10^{15}[1+0.5(1+\frac{n_{cr}(e)}{n(e)})]I([C\,II])\,\,\rm{cm^{-2}}.
\end{equation}

The relationship of the column density of ionized nitrogen to the integrated line intensity, in units of K \kmsno, in the optically thin limit, assuming uniform excitation along the line of sight, is given by,

\begin{equation}
I_{ul}([N\,II]) = \int T_{ul}(K)d{\rm v} =  \frac{hc^3}{8\pi k(\nu_{ul})^2}A_{ul}f_{u} N(\rm{N}^+)\,\rm{cm^{-2}},
\end{equation}

\noindent where $A_{ul}$ is the Einstein A-coefficient, $\nu_{ul}$ the transition frequency,  $f_{u}$ is the fractional population of the upper state which depends on the density of the collision partner, $n$, $I$ is in units of K \kmsno, and $N$(N$^+$) is the column  density of ionized nitrogen. For the $^3P_1 \rightarrow$ $^3P_0$ transition at 205 \micron this equation reduces to

\begin{equation}
I([N\,II]) = 5.06\times 10^{-16}  f_{1} N(\rm{N}^+)\,\,\rm{(K\,km\,s^{-1})}
\label{eqn:I_NII_N}
\end{equation}

\noindent where $f_1$ is the fractional population of N$^+$ in the $^3P_1$ state.  If we assume that the \nii emission region is uniform we can replace $N$(N$^+$) by, $n$(N$^+$)$L$, where $L$ is the path length in cm,

\begin{equation}
I([N\,II]) = 5.06\times 10^{-16}  f_{1} n(\rm{N}^+)L\,\,\rm{(K\,km\,s^{-1})}
\label{eqn:I_NII_xn}
\end{equation}

\begin{figure}
 \centering
      \includegraphics[width=7cm]{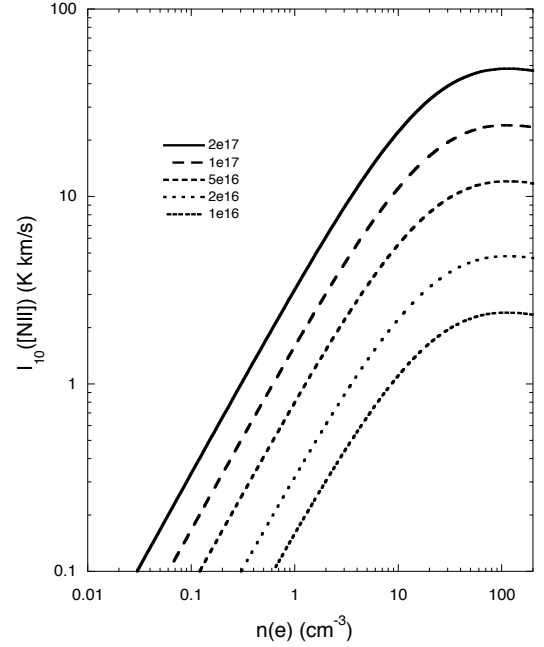}
      \caption{The intensity of \nii 205 \micron line versus $n$(e) for a range of ionized nitrogen column densities; the insert gives the key for the column density. The curves flatten out as the density approaches the critical density.}
              \label{fig:fig8}
 \end{figure}

We can solve for $f_1$  in the optically thin limit from the detailed balance equations for the $^3$P$_2$, $^3$P$_1$, and $^3$P$_0$ states along with the condition $f_0+f_1+f_2 =$1.  In ionized regions electrons dominate the excitation of N$^+$. The collisional de-excitation rate  coefficient by electrons is given by

\begin{equation}
C_{ul} = 8.63\times 10^{-6}\frac{G(u,l)}{g_u T^{0.5}},
\end{equation}

\noindent where $g_u$ is the degeneracy of the upper level and $G(u,l)$ is the temperature dependent effective collision strength \cite[see][]{Lennon1994,Hudson2004}. We calculated the electron collisional rate coefficients using the \cite{Hudson2004} values for the effective collision strengths.  The critical density for the 1-0 transition, where collisions out of the $^3P_1$ state balance the radiative transition, A$_{10}$, is $n_{cr}$(e) $\sim$ 44 cm$^{-3}$ at 8000K.  In Figure~\ref{fig:fig8}  we plot the intensity of the 205 \micron line as a function of the electron density using Equation~\ref{eqn:I_NII_N} for a range of ionized nitrogen column densities, $N$(N$^+$) = 10$^{16}$  to  10$^{18}$ cm$^{-2}$.  The intensity scales linearly with  $n$(e) up to $ \sim$10 - 15 cm$^{-3}$ beyond which it levels off as $n$(e) approaches the critical density. 

The \nii emission comes from ionized gas and so can be used to derive the electron density directly if we know the size of the emitting region and the fractional abundance of ionized nitrogen, $x$(N$^+$) = $n$(N$^+$)/$n_t$, where $n_t $= $n$(H)+2$n$(H$_2$)+$n$(H$^+$) is the total density of hydrogen nuclei including ions.  To estimate $n($e) we replace the  density of ionized nitrogen, $n$(N$^+$), in Equation~\ref{eqn:I_NII_xn} by $x$(N$^+$)$n_t$, and because the \nii emission arises from highly, or perhaps completely, ionized gas we set $n_t  \simeq n$(H$^+$) = $n$(e).  For a uniform density region we can take $N$(e) = $n$(e)$L$, where $L$ is the thickness of the emission region, and can now express Equation~\ref{eqn:I_NII_xn}  as

\begin{equation}
I([N\,II])=\int T_{10} dv = 5.06\times 10^{-16} x(N^+) L n(e) f_1(n(e)),
\end{equation}

\noindent where $I$ is in K \kmsno. Rewriting in convenient units we obtain, 

\begin{equation}
I([N\,II])=0.156x_{-4}(N^+) L_{pc} n(e)f_1(n(e)),
\end{equation} 

\noindent where $x_{-4}$ is the fractional abundance of N$^+$ is in units of 10$^{-4}$ and $L_{pc}$ is in pc.  Thus we can solve for the electron density, $n$(e), as a function of the observed intensity of the 205 \micron line, given the  size of the emission region and the fractional abundance of N$^+$.  At low densities, where $n$(e)$< n_{cr}$(e), it can be shown that 

\begin{equation}
n(e) \propto \left [\frac{I(N\,II)}{L_{pc}x_{-4}(\rm{N^+})} \right]^{\alpha}\,\,\rm{cm^{-3}}
\end{equation}

\noindent  where $0.5< \alpha <0.7$, for $n$(e)$<n_{cr}$(e). In the limit $n$(e)$<< n_{cr}$(e), such as found in the Warm Ionized Medium in the disk, $f_1 \sim n$(e)/$n_{cr}$(e) and $\alpha =$0.5.   Thus, in general, $n$(e) is only moderately sensitive to uncertainties in $L$ and $x$(N$^+$).
 
 \subsection{The -207 \kms component}

As seen in Figures~\ref{fig:fig4} and~\ref{fig:fig5} there is \cii emission along the line of sight to the CO molecular cloud with V$_{lsr} \sim$ -207 \kmsno.  The peak in \cii intensity for this component is at $b>$ 0\deg at the edge of the lowest contour of CO and is thus consistent with limb brightened \cii emission.  Therefore the SOFIA \cii and \nii spectra along $b$=0\deg likely arise from a layer of gas whose size is related to the thickness of the \cii emission beyond the lowest CO contour.  The thickness of the \cii emission is about 6\arcmin which corresponds to  $\sim$15 pc at the distance to the CMZ.  To solve for the electron abundance in the \nii regions we need, in addition to $L_{pc}$ and $I($\niino$)$,  $x(\rm{N}^+)$ appropriate to the CMZ.  To estimate the fraction of N$^+$ in the CMZ we use the nitrogen abundance Galactic gradient, 0.07 dex kpc$^{-1}$, for $R_G >$ 3 kpc, derived from \cite{Rolleston2000} which leads to $x($N$^+$) $\propto exp(-R_G/6.2)$ \cite[see][]{Pineda2013,Langer2014}.  However because we do not have information about how it scales to smaller $R_G$ we extrapolate this relationship to the half way point, 1.5 kpc.   Using the local nitrogen abundance fraction $x(N^+)$ =5.51$\times$10$^{-5}$ from \cite{Jensen2007} yields $x$(N$^+$) =  1.6$\times$10$^{-4}$ at R$_G$=1.5 kpc.    As discussed above, the solutions for $n$(e) are not very sensitive to modest uncertainties in $x$(N$^+$).  The solutions for $n$(e) and $N$(N$^+$) are given in Table~\ref{tab:Table2}. 

The electron density for the -207 \kms component varies from $\sim$9 to 21 cm$^{-3}$ for the four LOS with \nii detections and has an average value $\sim$14 cm$^{-3}$.  Thus the gas traced by \nii is warm dense ionized gas, significantly denser than that derived for the typical WIM in the disk, which has $n$(e) of order a few$\times$10$^{-2}$ cm$^{-3}$ \citep{Haffner2009,Velusamy2012}. In the next section we will discuss what might generate and sustain such a high density ionized gas towards the cloud associated with Sgr E.
 
\begin{table}[!htbp]																	
\caption{Electron density and nitrogen column density}
\label{tab:Table2}															
		\begin{tabular}{lcccccc}																	
\hline	
  V$_{lsr}$$^a$ =  &   -207  &  -207  & -207 &  -174  &  -174  & -174\\
  LOS &  $n$(e)$^b$ &  $N$(N$^+$)$^c$ &  $N$(H$^+$) & $n$(e)$^b$ &     $N$(N$^+$) & $N$(H$^+$)  \\
  \hline
  \hline 
358.45$^d$  &   20.7 &   1.5e17 & 9.3e20 & 9.9 &  7.5e16 & 4.6e20 \\
358.55  &  14.6 & 1.1e17 & 6.8e20  & 10.4 &   7.9e16 &  4.8e20 \\
358.60  &  9.1 & 6.9e16 & 4.2e20 & 6.3 & 4.8e16 &   2.9e20 \\
358.65  & - & -  &  - & - & - & -  \\
358.70  &  12.8 & 9.7e16 & 5.9e20 & 7.7 &  5.9e16 & 3.6e20  \\
358.75  & - & -  &  - & -  & - & - \\
\hline
Average &  14.3 & 1.1e17 & 6.6e20 & 8.6 &  6.5e16 & 4.0e20 \\
\hline
\hline		
\end{tabular}
\\
a) In \kmsno. b) Densities in cm$^{-3}$. c) Column densities in cm$^{-2}$ and rounded to one decimal place.  d) All LOS are along $b=$0\degno.
\end{table}

Assuming that the nitrogen is nearly fully ionized, the column density of ionized carbon in this region is given by $N$(C$^+$)=$x$(C$^+$)/$x$(N$^+$)$N$(N$^+$).  For  $x($C$^+$)/$x$(N$^+$) we adopt the value of 3.2 \citep{Jensen2007,Sofia2004} and the values for N(C$^+$) are given in Table~\ref{tab:Table3} along with $N$(H$^+$).  The C$^+$ column densities are of order a few$\times$10$^{17}$ cm$^{-2}$, with an average $\sim$4$\times$10$^{17}$ cm$^{-2}$.  
 
The \cii intensity associated with the \nii region, $I_{\rm{H}^+}$(\ciino), is calculated using the inverse of Equation (2), from the column density $N$(C$^+$), $n$(e) derived from \niino, and assuming T$_{k}$ = 8000 K.  The exact kinetic temperature is not critical as the column density is only weakly dependent on temperature through the collisional de-excitation rate coefficient, resulting in $n_{cr}$(e) $\sim$ 50$T_4$$^{0.37}$ cm$^{-3}$, where $T_4$ is the kinetic temperature in units of 10$^4$ K.   The results are given in Table~\ref{tab:Table3}. 

The \cii emission from the neutral PDR, $I_{\rm{H_2}}$(\ciino), is the difference between the observed total \cii intensity, $I$(\ciino), given in Table~\ref{tab:Table1}, and the intensity calculated from the \nii region, $I_{\rm{H^+}}$(\ciino) \citep{Langer2014}.  This quantity, $I_{\rm{H_2}}$(\ciino) = $I$(\ciino) - $I_{\rm{H^+}}$(\ciino) arises primarily from the CO--dark H$_2$ gas (also called the CO--dark gas)  and is given in Table~\ref{tab:Table3}. Negative or very small values of $I_{\rm{H_2}}$(\ciino) indicate that almost all the \cii emission arises from the dense ionized gas.  Unfortunately we do not know the kinetic temperature and hydrogen density of the CO--dark $N$(H$_2$) to calculate accurately its measured C$^+$ column density.  For illustrative purposes we calculate the column density of C$^+$ in the H$_2$ layer, $N_{\rm{H_2}}$(C$^+$), and the corresponding column density of CO--dark H$_2$, $N_{\rm C^+}$(H$_2$)= $N_{\rm{H_2}}$(C$^+$)/$x$(C$^+$) \cite[see][]{Langer2014},  assuming 100K and 300 cm$^{-3}$ in Equation (1), along with the appropriate $n_{cr}$(H$_2$) from \cite{Wiesenfeld2014}. These results are listed in Table~\ref{tab:Table3} for all the LOS with positive $I_{\rm{H_2}}$(\ciino).  Higher densities will result in lower column densities and vice versa. 

\begin{table}[!htbp]																	
\caption{C$^+$ column densities and intensities}
\label{tab:Table3}															
		\begin{tabular}{lccccc}																	
\hline	
 LOS$^a$   &   $N$(C$^+$)$^b$  & $I_{\rm{H^+}}$(\ciino)$^c$  & $I_{\rm{H_2}}$(\ciino) & $N_{\rm{H_2}}$(C$^+$) & $N_{\rm{C^+}}$(H$_2$)$^f$\\
  \hline
  \hline 
 -207$^d$  \\  
\hline
358.45 &  4.9e17   & 64.1  &  -0.7  & -  & -\\
358.55  &  3.5e17 & 38.5 &  7.0 &  4.2e17 & 8.2e20 \\
358.60 &  2.2e17   & 18.3 &   29.3  & 1.8e18 & 3.5e21 \\
358.65  & -  &  - & -   & - & -\\
358.70  & 3.0e17  & 31.3 &  14.1 & 8.6e17 & 1.7e21 \\
358.75  &  - & -  & -  & -  &- \\
\hline
Average & 3.4e17 & 38.1 &  16.8$^e$ & 1.0e18 & 2.0e21 \\
\hline
\hline
-174$^d$  \\ 
\hline
358.45  &  2.4e17  &  21.1  & 0.1 &  3.3e15 & 6.4e18 \\
358.55   & 2.5e17  & 22.6 & -10.4  &  - & - \\
358.60  &  1.5e17 &  9.9  & 33.1  &  2.0e18 & 3.9e21 \\
358.65  & - & - & -   & -  & - \\
358.70  &  1.8e17  & 13.4  & 1.3 & 8.2e16 & 1.6e20 \\
358.75  & - & -  &   - & -  & - \\
\hline
Average & 2.0e17 & 16.9 & 11.5$^e$ & 7.0e17 & 1.4e21 \\
\hline
\hline		
\end{tabular}
\\
a) All LOS are at $b=$0\degno.  b) Column density in cm$^{-2}$ and rounded to one decimal place. c)   Intensity of \cii in the \nii emission region in K \kmsno.	d) V$_{lsr}$ in \kmsno.  e) Negative intensities are not included. f) Assumes $T_k$=100 K and $n$(H$_2$) = 300 cm$^{-3}$.
\end{table}

\subsection{The -174 \kms component}

The -174 \kms \nii and \cii emissions are weaker, and in some positions much weaker, than those of the -207 \kms component, and are associated with CO only at $l=$358\fdg 75 in Figure~\ref{fig:fig6}. The CO maps in \cite{oka1998a} show a CO molecular cloud with an edge at this longitude and extending inwards to larger $l$. The -174 \kms emission is also associated with  \hi emission (Figure~\ref{fig:fig7}) that is weaker than that of the -210 \kms cloud. We do not detect this weaker \cii component at all the positions in the HIFI OTF maps (see discussion in Section 2.2) and so have no estimate of the size of the \cii and \nii emission region. For comparison we have calculated densities and the column densities of this component assuming the  same size, 15 pc, for the \nii emission region as that for the -207 \kms cloud.  These results are given in Tables~\ref{tab:Table2} and ~\ref{tab:Table3}.  The solutions for the -174 \kms component yield slightly lower average electron densities and column densities than for the -207 \kms component.



\section{Discussion}
\label{sec:discussion}

The analysis of \nii emission at the edge of the CMZ reveals a hot dense plasma with an electron abundance ranging from $\sim$ 6 to 21 cm$^{-3}$.  The densities we derive are much larger than those characteristic of the disk's WIM, but are consistent with those derived for very bright nebula with numerous luminous O-type stars. For example, \cite{Oberst2011} studied the Carina Nebula, which has a very large UV flux, using emission from the \nii 205 \micron and 122 \micron lines along with an excitation model\footnote{With two transitions the electron density can be derived without having to assume a characteristic size for the emission region.}, and find values of $n$(e) ranging from a few to over 100 cm$^{-3}$.  In the discussion below we will use the results from our analysis of \nii and \cii emission to characterize some of the possible energy sources that  can maintain such a hot dense ionized plasma. 
 
\subsection{Ionization of Nitrogen}

Nitrogen has an ionization potential of 14.534 eV, higher than that of hydrogen at 13.598 eV, and in the ISM can be ionized by cosmic rays, X-rays, Extreme Ultraviolet (EUV) photons with wavelengths shortward of 853.06 \AA  , electron collisional ionization, or via charge exchange with protons (H$^+$) at energies $\geq$0.936 eV, the energy difference between the ionization of H and N.  In a nearly fully ionized region N$^+$ is destroyed primarily via electron radiative and dielectronic recombination, and in partially ionized regions also via exothermic charge exchange with H atoms. We can estimate the required ionization rate to sustain  a highly ionized N$^+$ gas by calculating the electron recombination rate for the solutions for $n$(e) given in Table~\ref{tab:Table2}. The recombination rate coefficients are taken from the UMIST data base \citep{McElroy2013} and \cite{Badnell2006}  for radiative recombination and \cite{Badnell2003} for dielectronic recombination. (In molecular hydrogen clouds N$^+$ reacts rapidly with H$_2$ and initiates a chain of reactions producing nitrogen bearing molecules, but this channel is not relevant in the WIM, WNM, or \hii regions.)   Charge exchange with H has a reaction rate coefficient at 8000 K $\sim$5.5$\times$10$^{-14}$ cm$^3$ s$^{-1}$ (see Table 2 of \cite{Lin2005}) and when compared with recombination with electrons $\sim$8$\times$10$^{-13}$  cm$^3$ s$^{-1}$, we see that electron recombination dominates for $n$(e)/$n$(H) $>$ 0.2.  To get an estimate of the degree of ionization needed to sustain a highly ionized gas we neglect the charge exchange of N$^+$ with H in the regime where the gas is highly ionized and consider only recombination by electrons. For large electron densities in the range 5 to 25 cm$^{-3}$ such as derived here, the recombination rate ranges from 4$\times$10$^{-12}$ to  2$\times$10$^{-11}$ s$^{-1}$.  Sustaining a high ionization fraction ($\ge$0.5) requires an ionization rate comparable to or greater than these values .  For typical WIM conditions in the disk, $n$(e)$\sim$10$^{-2}$ cm$^{-3}$, the ionization rate required to keep nitrogen highly ionized is two to three orders of magnitude lower.  In the following we discuss the likely sources of ionization at the edge of the CMZ. We exclude electron collisional ionization because, due to the high ionization potential of N, it requires a very hot electron gas with kinetic temperatures of order 10$^5$K, a condition met only in the HIM portion of the CMZ.

\subsubsection{Cosmic ray ionization}

\cite{Goto2014} have detected H$_3$$^+$ in the CMZ and use its abundance to derive a lower limit on the hydrogen ionization rate, $\zeta_H >$ 1$\times$10$^{-15}$ s$^{-1}$, which they suggest could come from cosmic rays or X-rays.   If the ionization is due to cosmic rays, it is about ten to thirty times larger than the cosmic ray ionization rate in the disk \citep{Indriolo2012}. The cosmic ray ionization rate of nitrogen (including the secondary photons generated by cosmic rays interacting with hydrogen) is about a factor of two larger than that of hydrogen \citep{McElroy2013}, $\sim$few$\times$10$^{-15}$ s$^{-1}$, but is much lower than the values needed to offset the recombination by electrons.  Although regions of intense cosmic ray ionization may exist with ionization rates $\sim$10$^{-14}$ s$^{-1}$  \citep{Bayet2011,Meijerink2011}, rates $\ge$10$^{-12}$ s$^{-1}$ would be required for cosmic rays to explain high fractional nitrogen ionization at densities $n$(e) $\ge$ 1 cm$^{-3}$. Indeed \cite{Meijerink2011} consider rates almost that large in their models of photon dominated regions (PDRs) but only consider high atomic hydrogen densities, $n(H)\ge$10$^3$ cm$^{-3}$ such that the fractional ionization is small. Their models cannot be applied to the ionized gas probed by our \nii observations\footnote{Note that  \cite{Meijerink2011} calculate the N$^+$ fractional abundance, $x$(N$^+$)=$n$(N$^+$)/$n$(H+2H$_2$), as $\le$10$^{-7}$, which is a low ionization scenario.  They also calculate the line intensities of the \nii 205 \micron and 122 \micron lines, but their predictions fall far below the measured intensities given in Table~\ref{tab:Table1}.}. 

\subsubsection{X-ray ionization}

The CMZ has several different sources of X-rays, including an accreting black hole at the center, a bright X-ray source 1E1740.7-2942 located $l \sim$ 359\fdg 1 and $b \sim$ -0\fdg 1 \cite[e.g.][]{Heindl1993,Wang2002}, extended diffuse X-ray emission \citep{Koyama2007}, as well as over 9000 X-ray point sources \citep{Muno2009} some of which lie at the edge of the CMZ between $l =$ 358\fdg 9 and 359\fdg 0 around $b\sim$ 0\deg and could provide the $>$10$^{-15}$ s$^{-1}$ ionization rate of hydrogen required to explain the H$_3$$^+$ observations of \cite{Goto2014}.  (Unfortunately, these  {\it Chandra} observations of the CMZ survey the Galactic Center do not extend into the Sgr E region.) However, unlike ionization by cosmic rays, the cross section for ionization of heavy atoms by X-rays can be much larger than hydrogen,  primarily due to K shell ionization at X-ray energies $>$0.4 keV, and thus X-rays could play an important role in ionizing nitrogen. For example, at 1 keV the ionization cross section of nitrogen is $\sim$7$\times$10$^{-20}$ cm$^{-2}$ \citep{Verner1993} over  three orders of magnitude larger than that for hydrogen \citep{Goto2014,Wilms2000}.  Therefore, if the X-ray ionization rate needed to explain the production of H$_3$$^+$ is $\sim$few$\times$10$^{-15}$ s$^{-1}$,  then the corresponding ionization rate for nitrogen will be of order a few$\times$10$^{-12}$ s$^{-1}$, sufficient to explain the observed large fractional ionization of nitrogen at $n$(e) $\sim$ 5 to 25 cm$^{-3}$. The penetration depth of 1 keV X-rays is a column density $N$(H+2H$_2$) $\sim$ 10$^{22}$ cm$^{-2}$, which is sufficient to penetrate the diffuse gas and envelopes of clouds in the CMZ \citep{Goto2014}.  Whether X-rays are the primary ionization source or not depends on the details of the distribution and luminosity of the X-ray sources, and we do not have sufficient information to make this determination. 
 
\subsubsection{Photoionization}
 
The photoionization cross section $\sigma$ for ionizing nitrogen ranges from about 10$^{-17}$ to 1.5$\times$10$^{-17}$ cm$^2$ from the threshold at $\sim$853 \AA \space to  500 \AA \space \citep{Samson1990}.  The total photoionization rate is given by $\int_{\lambda <853 \AA} \sigma (\lambda) \Phi(\lambda)d\lambda$, where $\Phi$ (=4$\pi J$) is the incident flux integrated in units of photons cm$^{-2}$ s$^{-1}$.  It has been estimated that to maintain the ionization of hydrogen in the WIM one needs an ionizing flux of Lyman continuum photons $>$10$^5$ photons cm$^{-2}$ s$^{-1}$ \citep{Haffner2009}. \cite{Reynolds1995} estimate an incident Lyman continuum flux 4$\pi$J $\sim$ 2$\times$10$^6$ to 9 $\times$10$^6$ photons cm$^{-2}$ s$^{-1}$, sufficient to explain the ionization in the WIM.  However nitrogen is ionized by EUV photons beyond the Lyman limit, for which the flux must be determined from models incorporating a distribution of O and B stars and attenuation by the surrounding ISM, including considerations of the clumpiness of the ISM \cite[cf.][]{Haffner2009}.  In general the flux drops sharply beyond the Lyman limit, as can be seen in the model of a spectrum of a star cluster in Figure 17 of \cite{Kaufman2006}, where the spectral flux (in ergs/s/Hz/sr/cm$^2$) drops by a factor of 10 at the Lyman edge and by another factor of 10 by $\sim$400 \AA.  We can estimate the flux needed to sustain a high nitrogen ionization fraction by balancing photoionization with electron recombination,

\begin{equation}
n(N^+)/n(N^0)= \frac{\int_{\lambda<853 \AA} \phi (\lambda) \sigma (\lambda) d\lambda}{k_r(e)n(e)}
\end{equation} 

\noindent where $k_r$(e) is the electron recombination rate coefficient, $n$(N$^0$) is the neutral nitrogen number density, and $\phi$ the photon flux as a function of wavelength.  This equation can be simplified by assuming that the ionization cross section is roughly constant over the energy range of interest (see above), so $n$(N$^+$)/$n$(N$^0$) $\sim$ ($\Phi \sigma$)/($k_rn($e)), where $\Phi$ is the total photon flux in cm$^2$ s$^{-1}$. Therefore the fraction of ionized nitrogen is

\begin{equation}
f(N^+)=\frac{\Phi \sigma}{k_rn(e) + \Phi \sigma}
\end{equation}

\noindent For example, to sustain $f$(N$^+$) = 0.5 in dense ionized gas with T=8000 K at $n$(e) = 1, 10, and 20 cm$^{-3}$ requires an EUV flux $\sim$6.5$\times$10$^4$,  6.5$\times$10$^5$, and 1.3$\times$10$^6$ photons cm$^{-2}$ s$^{-1}$, respectively.  For a nearly fully ionized nitrogen regime the required intensity of the EUV field increases sharply.  For example, to maintain $f$(N$^+$) = 0.9 requires an order of magnitude greater flux than for $f$(N$^+$) = 0.5, which can be found near young massive stellar clusters \citep{Kaufman2006}. \cite{Rodriguez2004} estimate that the PDRs in the CMZ are illuminated by a far-UV radiation field about a factor of 10$^3$ larger than the local ISM, and if this increase extends to the EUV it would be sufficient to explain the ionization of nitrogen.  One constraint on the flux distribution near the -207 \kms Sgr E feature comes from observations of carbon and helium recombination lines.  \cite{Wenger2013} detected carbon recombination lines in two of the eight Sgr E \hii sources detected in hydrogen recombination lines, but did not detect any helium recombination lines.  This result indicates that the flux at wavelengths short of 504.26\AA  \space (energies $\ge$ 24.59 eV) is small.

\subsubsection{Proton charge exchange ionization}

The last important ionization mechanism is charge exchange with energetic protons.  In a highly ionized hydrogen gas, $n$(H$^+$) $>$ 50 $n$(H), we can neglect the neutralization of N$^+$ by charge exchange with H atoms, which yields the ratio of ionized to neutral nitrogen,

\begin{equation}
f(N^+) = n(N^+)/n(N^0) = \frac{k_x(T) n(H^+)}{k_rn(e)},
\end{equation}   

\noindent where the charge exchange reaction rate coefficient, $k_{x}(T)$ is a Maxwellian average of the cross section over the electron velocity distribution \citep{Lin2005}. The value of $k_{x}(T)$ is an exponential function of temperature because only protons with energies $\gtrsim$0.94 eV can undergo charge exchange. To a very good approximation, $n$(e) $\simeq n$(H$^+$) and the fractional abundance of nitrogen is

\begin{equation}
f(N^+) = \frac{k_x(T)}{k_x(T) + k_r(T)},
\label{eqn:CX_solution}
\end{equation}

\noindent which is {\it independent} of $n$(e) and $n$(H$^+$). However, the fraction $f$(N$^+$) is very sensitive to the temperature of the plasma because of the endothermic nature of the reaction. The charge exchange process for N$^+ $+ H $\rightarrow$ N + H$^+$ has been measured (Stebbings et al. 1960), but only above $\sim$30 eV, and the low energy cross sections must be calculated theoretically.  \cite{Lin2005} have made the most recent calculations of the charge transfer reaction with ground state nitrogen, H$^+$ + N $\leftrightarrow$ H + N$^+$, and use these to calculate the rate coefficients, $k_x$ for forward and reverse reactions. The results for the charge exchange reaction rate coefficients are given in their Table 2 and can be used to calculate the fractional nitrogen ionization as a function of kinetic temperature.   

\begin{figure}
 \centering
      \includegraphics[width=7cm]{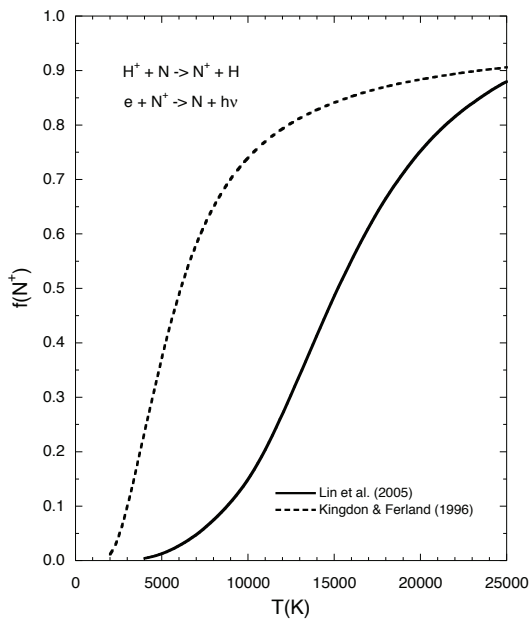}
      \caption{The fractional nitrogen abundance as a function of kinetic temperature due to production by proton charge exchange and destruction by electron recombination.  Two charge exchange reaction rate coefficients are plotted; the solid line is from \cite{Lin2005} and the dashed line from \cite{Kingdon1996}.}
              \label{fig:fig9}
 \end{figure}

However, one must be cautious here, because the theoretical rate coefficients derived by  \cite{Lin2005} for H$^+$ + N $\rightarrow$ N$^+$ + H  are lower than those of \cite{Steigman1971,Butler1979,Kingdon1996} (see Figure 7 in \cite{Lin2005}) below 2$\times$10$^4$ K, which is the relevant temperature regime for the WIM and \hii regions.  We solved Equation~\ref{eqn:CX_solution} for the rates derived by \cite{Lin2005} and \cite{Kingdon1996} as a function of temperature and these are plotted in Figure~\ref{fig:fig9}.  Using the reaction rates from \cite{Kingdon1996} leads to a high fractional abundance at temperatures consistent with those measured in the WIM and \hii regions, for temperatures 4000 to 10000 K.  However the rate coefficients from \cite{Lin2005} lead to a much lower ionization fraction of nitrogen at these temperatures, and requires temperatures $\gtrsim$15,000 K to ionize nitrogen efficiently via H$^+$ charge exchange.  Thus, depending on which calculation of $k_x$ is correct determines whether or not charge exchange with  protons can sustain a high ionization fraction of N$^+$ in the \nii region.

 \subsection{Conclusions}
    
From our analysis of \nii emission at the edge of the CMZ near Sgr E we find evidence for dense, hot, highly ionized gas. The association with \cii and CO indicates that this dense ionized gas surrounds the PDRs associated with two molecular clouds, one with V$_{lsr}$ $\sim$ -207 \kms and the other $\sim$ -174 \kmsno.  The electron densities, $n$(e) $\sim$ 5 to 21 cm$^{-3}$, are considerably higher  than those in the WIM of the disk, but consistent with those expected for bright diffuse \hii regions, such as the Carina nebula \citep{Oberst2011}. The electron densities  are considerably lower than those suggested by \cite{Lazio1998}, $n$(e)$\gtrsim$ 10$^3$ cm$^{-3}$,  if the photoionized layers of clouds are responsible for the scattering of radio waves in the CMZ. However, they are consistent with their other suggestion that the scattering could be due to the interface with a hot ionized medium with $n$(e)$\sim$ 5 - 50 cm$^{-3}$ and $T_{k}$(e)$\sim$ 10$^5$ -- 10$^6$ K.  We have no information about the kinetic temperature from our observations, and have assumed typical WIM kinetic temperatures $\sim$ 8000K.  Furthermore, our results apply to the ionized regions around the molecular cloud and if these are indeed the source of the scattering of radio waves, then they occupy a small filling factor in the CMZ and may not be representative of the electron density in the bulk of the volume.

There are three ways to sustain such a dense highly ionized hydrogen gas with large nitrogen ionization fraction.  One is to have a strong EUV radiation field with a flux of order 10$^6$ to 10$^7$ photons cm$^{-2}$ s$^{-1}$, another is to have a high enough kinetic temperature for rapid proton charge exchange with nitrogen, and the third is if there is a flux of X-rays sufficient to ionize nitrogen in the diffuse gas.   In the case of photoionization and X-ray ionization it is not necessary for the gas to be very hot.  For charge exchange, a gas temperature of order 5000 K suffices for the rates from \cite{Kingdon1996}, but a much higher temperature of order 15000 K is required if the rates calculated by \cite{Lin2005} are correct.  We are not able to distinguish which one of these processes dominates the ionization of the gas observed in \niino, or whether all three contribute.  The asymmetry in the \cii limb brightening in $b > $0\deg versus $b <$ 0\deg may be indicative of an asymmetry in the distribution of ionizing sources.  However we lack a complete survey of the Sgr E region in X-rays or EUV from which to draw any firm conclusions. 

The edge of the CMZ extends a little farther outwards (smaller $l$) in ionized components than in CO.  The ThrUUMS CO map,  Figure~\ref{fig:fig6}, detects CO at -207 \kms out to $l=$ 358\fdg 50, whereas the GREAT  \cii and \nii observations show emission out to at least 358\fdg 45, a difference of 7 pc.  Whether the CMZ extends any further cannot be determined as the GREAT observations did not extend to smaller values of $l$, and the HIFI OTF maps, which extend to 358\fdg 20, are less sensitive than the GREAT \cii data and show only a hint of \cii at -207 \kms extending to 358\fdg 35, a distance of 14 pc from the CO edge.

We have also estimated the column density of CO--dark H$_2$ from \cii observations and find an average column density of order 1-2$\times$10$^{21}$ cm$^{-2}$.  Models of massive molecular clouds \citep{Wolfire2010,Bolatto2013} predict  such column densities, somewhat independent of the mass of the cloud and of the intensity of the UV field.  Observations of \cii and CO towards a large ensemble of molecular clouds in the Galactic disk \citep{Langer2014}  are consistent with this model. Thus clouds in the CMZ and disk have similar CO--dark H$_2$ layers despite differences in UV radiation field and density.



\section{Summary}
\label{sec:summary}

We have explored the ionized gas at the edge of the CMZ near Sgr E, combining deep integrations of spectrally resolved \cii (158$\mu$m) and \nii (205$\mu$m) at six lines of sight along $b$ = 0\deg made with the GREAT instrument on SOFIA and HIFI spectrally resolved \cii on-the-fly strip maps in $b$ at thirteen longitudes spaced 0\fdg 05 apart.  We detect two distinct features in the \cii and \nii data, one on the line of sight to a CO molecular cloud at V$_{lsr} \sim$ -207 \kmsno, associated with Sgr E, and the other at $\sim$ -174 \kms outside the edge of another CO cloud.

We find that the brightest emission in the HIFI strip maps toward the cloud at -207 \kms forms an arc of \cii emission above the plane at the edge of the lowest CO contours.  It likely represents limb brightened emission in a highly ionized layer outside the molecular cloud.  We also detected \nii emission at a few positions in this edge, and both \cii and \nii have very broad lines there with $\Delta V$ of order 25 to 35 \kmsno.  The electron density in this region was derived from \nii emission and a radiative transfer model, assuming a characteristic size.  We find $n$(e) $\sim$ 9 to 21 cm$^{-3}$, about three orders of magnitude higher than that characterizing the disk's warm ionized medium (WIM), much smaller than those of compact \hii regions, but consistent with densities derived from excitation analysis of \nii in bright extended  \hii regions, such as the Carina nebula \citep{Oberst2011}.  The electron densities determined from the \nii for the ionized envelope of the cloud associated with Sgr E are lower than those suggested  to explain the scattering of radio waves used to derive the average electron density distribution in the CMZ if the scattering comes from the ionized envelope \citep{Lazio1998}. 

The ionization of such a dense region requires  a large flux of EUV radiation, and/or X-ray flux, and/or high enough temperatures to allow rapid charge exchange with protons via the endothermic reaction,  H$^+$ + N $\rightarrow$ H + N$^+$. The presence of abundant N$^+$ requires this region to be hot, and the heating could be photoelectron emission from the FUV, EUV, and X-ray radiation fields, dissipation of strong magnetic turbulence, or shock heating of supersonic turbulence.  This region has many compact bright \hii sources, indicating that it is an active massive star formation region, and such stars have the capacity to ionize and heat the gas in this region.  The CMZ is also a source of diffuse X-rays and  discrete X-ray sources that could provide the X-ray flux to ionize the nitrogen (as well as the hydrogen), but it depends on details of the distribution and luminosity of the X-rays.  Finally, we estimate that this CO cloud has a layer of CO--dark H$_2$ with column density $\sim$1-2$\times$10$^{21}$ cm$^{-2}$, consistent with theoretical models and observations of clouds in the disk.

We also detect ionized gas at a velocity $\sim$-174 \kms and the lines of sight observed in \cii and \nii with GREAT have \hi but no CO except except at $l =$ 358\fdg 75 which turns out to be the edge of a molecular cloud mapped in CO \citep{oka1998a}.   However this \cii emission is weak and the number of \cii detections in the HIFI map is insufficient to determine the size of  the ionized envelope for this cloud.  Therefore to estimate the properties of the envelope  we assumed the same thickness as that for the -207 \kms cloud.  An analysis of the \nii emission using parameters similar to those for the -207 \kms cloud suggests that it too is a hot high density ionized gas with $n$(e)$\sim$ 6 to 10 cm$^{-3}$.   The physical conditions in this ionized component are likely to be similar to those at the edge of the molecular cloud at $\sim$-207 \kms and have similar heating and ionizing sources.  Larger scale maps of spectrally resolved \cii and \nii will be needed to understand further the role of the ionized gas at the edge of the CMZ.   Finally, the CMZ is seen to extend further out in ionized gas from the CO edge by upwards of at least 7 to 14 pc.



\begin{acknowledgements}
This work is based in part on observations made with the NASA/DLR Stratospheric Observatory for Infrared Astronomy (SOFIA). SOFIA  is jointly operated by the Universities Space Research Association, Inc., under NASA contract NAS2-97001, and the Deutsches SOFIA
Institut (DSI) under DLR contract 50 OK 0901 to the University of Stuttgart. We would like to thank Drs. R. G$\ddot{\rm u}$sten and G. Sandell for their support  of  the observations and for keeping in close contact with us during the flights to adjust the observing plan as needed.  In addition we owe a special thanks to Dr. David Teyssier for clarifications regarding the {\it hebCorrection} tool.  We also thank an anonymous referee for comments and suggestions that improved the discussion in our paper. This work was performed at the Jet Propulsion Laboratory, California Institute of Technology, under contract with the National Aeronautics and Space Administration.   
 
\end{acknowledgements}



\bibliographystyle{aa}
\bibliography{aa_2014_25360_Langer_refs}


\end{document}